\documentclass[]{spie}  

 \usepackage[inline]{enumitem}

\usepackage{amsmath,amsfonts,amssymb}
\usepackage{graphicx}
\usepackage[utf8]{inputenc}

\graphicspath{ {./figures/} }
\usepackage[colorlinks=true, allcolors=blue]{hyperref}

\title{Intra-operative Brain Tumor Detection with\\Deep Learning-Optimized Hyperspectral Imaging}

\author[a]{Tommaso~Giannantonio}
\author[b]{Anna~Alperovich}
\author[a,c]{Piercosimo~Semeraro}
\author[c,d]{Manfredo~Atzori}
\author[e]{Xiaohan~Zhang}
\author[e]{Christoph~Hauger}
\author[f]{Alexander~Freytag}
\author[g]{Siri~Luthman}
\author[g]{Roeland~Vandebriel}
\author[g]{Murali~Jayapala}
\author[h]{Lien~Solie}
\author[h]{Steven~de~Vleeschouwer}

\affil[a]{Carl Zeiss AG, Corporate Research \& Technology, Kistlerhofstr. 70, 81379 Munich, Germany}
\affil[b]{Carl Zeiss AG, Corporate Research \& Technology, Carl-Zeiss-Str. 22, 73447 Oberkochen, Germany}
\affil[c]{University of Padua, Department of Neuroscience, Via Belzoni 160, 35121 Padua, Italy}
\affil[d]{Information Systems Institute, University of Applied Sciences Western Switzerland (HES-SO Valais), 3960 Sierre, Switzerland}
\affil[e]{Carl Zeiss Meditec AG, Rudolf-Eber-Str. 11, 73447 Oberkochen, Germany
}
\affil[f]{Carl Zeiss AG, Corporate Research \& Technology, Carl-Zeiß-Promenade, 07745 Jena, Germany}
\affil[g]{IMEC, Kapeldreef 75, 3001 Leuven, Belgium}
\affil[h]{Dept. Neurosurgery, University Hospitals Leuven \& Dept. Neuroscience, Leuven Brain Institute (LBI), Laboratory of Experimental Neurosurgery and Neuroanatomy, KU Leuven, Herestraat 49, 3000 Leuven, Belgium}

\authorinfo{Further author information: (Send correspondence to T.G.)\\T.G.: E-mail: name dot surname at zeiss dot com}

\begin{document} 
\maketitle

\begin{abstract}
Surgery for gliomas (intrinsic brain tumors), especially when low-grade, is challenging due to the infiltrative nature of the lesion. Currently, no real-time, intra-operative, label-free and wide-field tool is available to assist and guide the surgeon to find the relevant demarcations for these tumors. While marker-based methods exist for the high-grade glioma case, there is no convenient solution available for the low-grade case; thus, marker-free optical techniques represent an attractive option. Although RGB imaging is a standard tool in surgical microscopes, it does not contain sufficient information for tissue differentiation. We leverage the richer information from hyperspectral imaging (HSI), acquired with a snapscan camera in the $468-787\,$nm range, coupled to a surgical microscope, to build a deep-learning-based diagnostic tool for cancer resection with potential for intra-operative guidance. However, the main limitation of the HSI snapscan camera is the image acquisition time, limiting its widespread deployment in the operation theater. Here, we investigate the effect of HSI channel reduction and pre-selection
to scope the design space for the development of cheaper and faster sensors. Neural networks are used to identify the most important spectral channels for tumor tissue differentiation, optimizing the trade-off between the number of channels and precision to enable real-time intra-surgical application. We evaluate the performance of our method 
on a clinical dataset that was acquired during surgery on five patients. By demonstrating the possibility to efficiently detect low-grade glioma, these results can lead to better cancer resection demarcations, potentially improving treatment effectiveness and patient outcome.
\end{abstract}


\section{Introduction}
\label{sec:intro}

Brain tumors are among the most frequent tumor types worldwide with a high mortality. The most common type of brain tumors are gliomas, which are classified into high-grade gliomas (HGG, grade III and IV) and low-grade gliomas (LGG, grade I and II)\cite{lo2016t}. Among others, neurosurgical resection of gliomas still represents the primary treatment method but remains challenging due to their indefinite tumor margin under white light, as a result of their heterogeneity and infiltrative growth into the surrounding brain tissue\cite{JAKOLA20171942, pallud2010natural}. Consequently, tumor tissues can be resected incompletely, leading to a probable recurrence, which is a major cause of mortality \cite{sanai2008glioma}.  On the contrary, a larger safety margin might result in over-resection, i.e. removal of healthy brain tissue, resulting in permanent brain function damage\cite{stummer2011counter}.

Recently, various techniques became available to neurosurgeons for improved intra-operative visualization of gliomas, such as neuronavigation, intra-operative MRI and ultrasound \cite{leroy2019high, b2020intra}. Apart from these techniques, fluorescence-guided surgery using 5-aminolevulinic acid (5-ALA) induced fluorescence has become one of the most powerful methods for visualizing HGGs\cite{stummer1998intra, stummer1998tech,STUMMER2006392}. Nevertheless, the efficacy of this method was reported to be still unclear for resections of LGGs due to their limited fluorescence response\cite{almekkawi2020t}.  To the best of our knowledge, there are no fluorescence tracers available marking LGGs till now. It is thus of our interest to develop a real-time, wide-field technique that can improve visualization of LGG intra-operatively without the use of a tracer.

In this work, we propose a novel approach for intra-operative tissue differentiation in neurosurgery. Our solution utilizes hyperspectral image information in order to classify areas with healthy tissue and LGG, which is more challenging than the case of HGG and no previous attempt of which has been made to the best of our knowledge.

We illustrate that information about the tissue type is available in the spectral curves, and can be successfully extracted with a neural network. Furthermore, we perform model explainability and channel selection to investigate what spectral regions and channels carry the most discriminative power. This is important both for trusting the model outputs as well as to design less expensive and faster data acquisition strategies based on a reduced spectral sampling. Finally, we evaluate the predictions' reliability with an ensemble method \cite{ensemble} by generating a map of reliable predictions to support the doctor during surgery.

The rest of this paper is structured as follows: in Sec.~\ref{sec:related_work} we summarize previous related work;
in Sec.~\ref{sec:data} we describe the hardware, data acquisition process and resulting dataset. In Sec.~\ref{sec:methods} we describe our analysis methods, and in Sec.~\ref{sec:results} we summarize our results before concluding in Sec.~\ref{sec:conclusions}.

\section{Related work}
\label{sec:related_work}

\subsection{Hyperspectral imaging}
Hyperspectral imaging (HSI) \cite{kamruzzaman2016introduction} is a technology allowing the acquisition of images in many (dozens to hundreds) narrow spectral bands. As such it combines the main advantages of traditional imaging (description of morphological features) and spectroscopy (sensitivity to chemical composition).
Besides medicine, HSI has been successfully used in multiple fields such as remote sensing for land cover classification \cite{vali2020deep}, food quality control \cite{ozdougan2021rapid}, agriculture, \cite{lu2020hyperspectral} garbage sorting \cite{tatzer2005industrial}, astronomy \cite{benitez2014j, serrano2022pau}, art conservation \cite{cucci2016reflectance}, military \cite{shimoni2019hypersectral}.

\subsection{HSI in medicine}
Biological tissues have distinct spectral characteristics driven by their chemical composition \cite{jacques2013optical}. In the range $450 - 600$ nm blood is the main  component, dominated by hemoglobin (Hb). The Hb spectral shape varies depending on the oxygen saturation level, presenting a single absorption peak at 560 nm when deoxygenated and a double peak at 540 and 580 nm when oxygenated \cite{eaton1978optical}. 
As this and other parameters driving the spectral shape in absorption, scattering, and fluorescence differ between tissues and pathological states, HSI can be used to discriminate between them.

The use of HSI in medicine has been growing in the last decade\cite{lu2014medical, fei2020hyperspectral, ortega2020information}, and it is establishing itself as a non-invasive, non-ionizing, label-free diagnostic tool. 
As such, it has seen significant applications in many medical domains, including the analysis of multiple types of cancer (\emph{in vivo} and \emph{ex vivo}) \cite{halicek2019vivo}, computational pathology \cite{ortega2020hyperspectralREV}, gastroenterology \cite{ortega2019use}, dermatology \cite{johansen2020recent}, general surgery \cite{clancy2020surgical, shapey2019intraoperative, barberio2021intraoperative} (all references here are to recent review articles).

One of the most important aspects underlying all the above-mentioned studies is the quest for the optimal method to extract information from biomedical HSI data \cite{ortega2020information}.

The simplest method is based on optical inverse modelling: if a physical model for the tissue spectra exists, observations can be used to directly infer its parameters \cite{milanic2015detection}; e.g., the ratio of the spectral bands corresponding to oxygenated and de-oxygenated Hb can be used for cancer detection \cite{leon2022hyperspectral}.
In most cases however, an accurate physical model is missing, and a data-driven machine learning approach is more suitable, either of the classical feature learning (FL) type, or based on deep learning (DL).

Common classical algorithms used for pixel-wise classification in the medical HSI domain  include support vector machines (SVM), random forests (RF), and multinomial logistic regression (MLR) \cite{ghamisi2017advanced}.
Dimensionality reduction and feature selection methods are often applied to HSI data to discard irrelevant information, reduce computational time, and as an instrument to design custom HSI cameras that only acquire important information.

Beyond pixel-wise analyses, it is possible to consider the whole HS cube and employ spectra-spatial techniques \cite{he2017recent} to address the common issues of low inter-class and high intra-class (patient-to-patient) variability. Different strategies for the spectral-spatial integration exist at the pre-/post-processing, or integrated levels.

In recent years, the number of deep learning applications to the medical HSI domain has been increasing \cite{li2019deep}.
Several dimensionality approaches are possible: pixel-wise 1D convolutional neural networks (CNN); 2D CNNs applied in parallel to all spectral channel and then concatenated; or fully 3D CNNs.
The most popular approach has been 2D \cite{fabelo2019deep, ortega2020hyperspectral2, halicek2019hyperspectral, halicek2020tumor}, but some studies achieved improved results with 3D methods \cite{halicek2018tumor, manni2020hyperspectral}.

While most studies implemented (macro-)pixel-based classification or regression tasks, full-image segmentation tasks with 2D U-net architectures have also been addressed \cite{trajanovski2019tumor}.
Finally, advanced DL methods have been applied to HSI data, such as generative adversarial networks (GANs) to generate HSI from RGB data \cite{halicek2020conditional} and recurrent neural networks (RNN) to include the temporal element for video-based real-time inference \cite{bengs2020spectral}.

The main bottleneck that hampers a more widespread and successful use of DL in this domain is the scarcity and high cost of training data.

\subsection{HSI in brain cancer surgery}

We focus here on the applications to brain surgery \cite{wu2022review}.
A first HSI application to this field is to infer the brain tissue metabolic and hemodynamic signals, such as oxyhemoglobin, deoxyhemoglobin, and cytochrome c-oxidase, to study the functionality of the brain, diagnosing diseases, and for surgical assistance \cite{mori2014intraoperative, pichette2016intraoperative, giannoni2021hyperspectral, caredda2020optimal, iwaki2021novel}. Our main focus here is however the use of HSI for tumor tissue identification. For malignant cases such as gliomas, surgery is often the best treatment option, but the detection of the tumor edges is challenging with the naked eye. \cite{sanai2011extent} Existing intra-operative navigation tools, such as magnetic resonance imaging, ultrasound, or fluorescent markers, have significant limitations, so that HSI-based margin delineation is an attractive option.\cite{halicek2019vivo}

The state of the art for the \emph{in vivo} human brain cancer classification application is represented by the European project \emph{HELICoiD} (HypErspectraL Imaging Cancer Detection).\cite{fabelo2019deep, kabwama2016intra, fabelo2018spatio, fabelo2019surgical, ravi2017manifold, fabelo2018intraoperative, pineiro2017p04, fabelo2016helicoid}. These authors developed an intra-operative demonstrator to acquire and process HSI in real time in order to support the operating surgeon during resection \cite{fabelo2018intraoperative}, acquiring data in the VIS and VNIR ranges with high spectral resolution.
First tumor classification results from 33 HS cubes and 22 patients obtained specificity and sensitivity $>96\%$ using classical and deep learning methods \cite{fabelo2016helicoid}. 
Later work incorporated further data (36 cubes from 22 patients) and introduced a semi-automatic labelling tool to improve annotation quality; this database is publicly available \cite{fabelo2019vivo}. The results based on this dataset and a combination of classical and deep learning methods achieved sensitivity and specificity $> 98\%$ \cite{fabelo2018spatio}. The method was later further improved and tested on a fully functional demonstrator \cite{fabelo2018intraoperative} that could classify four tissue types. Later work revised the model implementation, to improve its parallelization \cite{florimbi2018accelerating} and achieve real-time processing on multiple GPUs \cite{florimbi2020towards, torti2018acceleration}.
Subsequent studies employing a deep learning-based pipeline further improved the accuracy by $\sim 16\%$ \cite{fabelo2019deep,fabelo2019surgical} with respect to classical methods, while requiring a larger amount of training data. 
The results were subsequently further improved with more advanced deep learning architectures: Ref.~\citeonline{manni2020hyperspectral} introduced a 3D-2D hybrid CNN, while Ref.~\citeonline{hao2021fusing} employed a multiple model fusion. Most recently Ref.~\citeonline{leon2021vnir} developed the fusion of VIS+NIR data, obtaining a 21\% improvement in the classification.

Several efforts were undertaken towards dimensionality reduction and suppression of redundant information, such as Fixed Reference $t$-distributed Stochastic Neighbors (FR-t-SNE) \cite{ravi2017manifold}, and methods to identify the most important spectral bands for classification based on classical feature selection methods such as the genetic algorithm, particle swarm and ant colony optimizations \cite{martinez2019most} and empirical mode decomposition \cite{baig2021empirical}.

As a separate part of the \emph{HELICoiD} project, an \emph{in vitro} histology dataset was also produced \cite{ortega2016hyperspectral} and employed for classification, achieving high-accuracy results with classical and later deep learning methods, \cite{ortega2018detecting, ortega2020hyperspectral2, ortega2020hyperspectral3} including using superpixel aggregation \cite{ortega2020hyperspectral}.
The experience gathered by the \emph{HELICoiD} group also led to the application of similar methods for the classification of skin cancers \cite{la2022neural, la2022deep}, Alzheimer's disease \cite{fabelo2020novel}, gastroenterology \cite{ortega2019use},  thyroid \cite{halicek2020tumor} and ENT cancers. \cite{halicek2019hyperspectral, halicek2019hyperspectral2, halicek2018tumor}.

An independent brain cancer dataset containing 13 images of 12 patients was collected and analyzed by Ref.~\citeonline{urbanos2021supervised} using a simpler HSI snapshot mosaic camera with 25 spectral channels only. These authors performed pixel-wise classification using classical and deep learning models, achieving good overall accuracy (95\%) when all patients are used in training, but limited generalization to unseen patients.

\begin{figure}[t]
    \centering
    \includegraphics[trim={0 0 0 0cm},clip,width=0.3\textwidth]{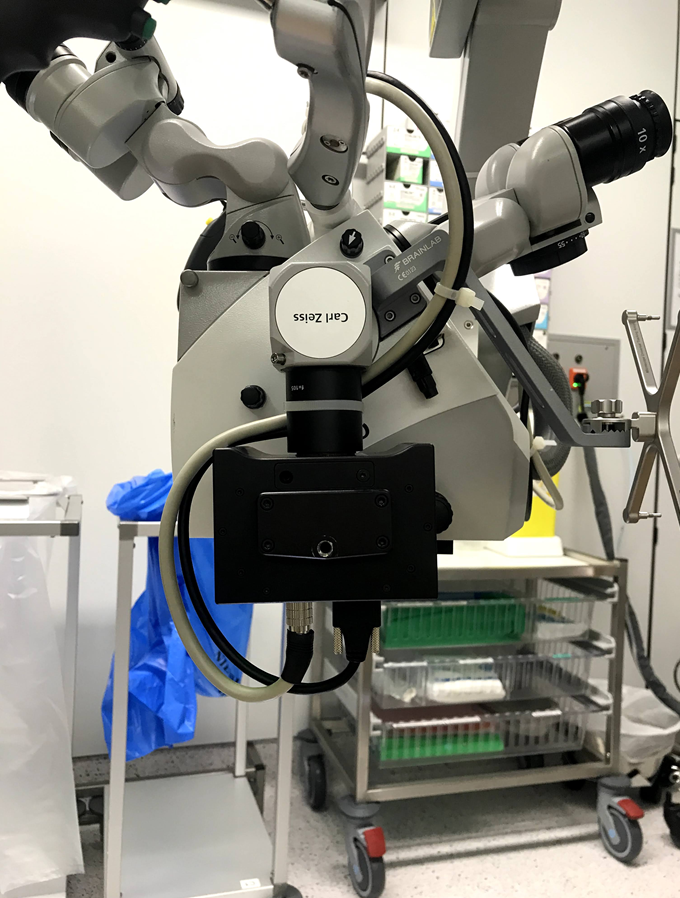}
    \caption{A photo of the setup in the operating room (OR) at UZ Leuven: IMEC HSI snapscan camera coupled to a ZEISS OPMI\textsuperscript{®} PENTERO\textsuperscript{®} microscope for intra-operative HSI and RGB data acquisition.}
    \label{fig:setup}
\end{figure}

\section{Hardware and data acquisition}
\label{sec:data}

The data acquisition was carried out intra-operatively in UZ Leuven using an IMEC snapscan VNIR (IMEC, Leuven, Belgium) hyperspectral camera, which was coupled to a ZEISS OPMI\textsuperscript{®} PENTERO\textsuperscript{®} 900 surgical microscope (Carl Zeiss Meditec AG, Oberkochen, Germany) at its documentation optical port through a video adapter with C-mount interface ($f=105$ mm) (see Fig.~\ref{fig:setup}).
 For each included patient, altogether 3 study HSI images were acquired, namely a first image of the brain surface after opening the dura sheet, a second image after partial removal of the tumor and a third image of the cavity after resection. To obtain absolute reflectance values, a Zenith Polymer\textsuperscript{®} target with 95\% reflectance was scanned by the HSI camera prior to the operation. Immediately prior to each HSI image acquisition, an RGB photo was taken using the PENTERO microscope, without varying the light condition nor the region of interest. For both RGB and HSI acquisitions, the internal illumination of a xenon light source (300 W) was used, and the operating room was darkened. For all image acquisitions, a working distance of 250 mm of the PENTERO microscope was used and the light intensity was set to 50\%. All raw HSI images were acquired with the IR800 mode enabled and were subsequently normalized and processed using the HSI snapscan software. \cite{pichette2017fast,pichette2017hyper}
The IMEC snapscan HSI camera delivers over 150 spectral bands in the range of $470-900$ nm. Nevertheless, due to the limitation caused by the IR cut filter in the illumination light path of the PENTERO microscope, we modified the calibration file of the HSI camera such that it covers a scan range of $470-780$ nm, resulting in 104 spectral bands. A detailed description of the setup and data acquisition workflow can be found in a companion work by some of us.\cite{Roeland2023integrating}

The acquired RGB images were annotated by the operating surgeon directly after the respective surgeries into classes that will be further discussed in Sec.~\ref{sec:data_explore}. The annotations were then registered to the HSI images accordingly to compensate the slight mismatch between the FOV of the HSI camera and the internal RGB camera of the PENTERO microscope, based on registration parameters determined using a spatial reference target printed on an A4 paper.

\section{Analysis methods}
\label{sec:methods}

We approach the problem of tissue differentiation in several steps. First, we analyze the data and confirm the assumption that for good quality data the healthy tissue has a different spectrum than a tumor. Then we investigate the discriminative capabilities of  classical Machine Learning (ML) and Deep Learning (DL) techniques. We perform an explainability analysis to identify the most important spectral channels for the task of  tissue differentiation and suggest an uncertainty estimation method to improve the robustness of the neural network predictions provided to the surgeon.

\subsection{Data exploration and preparation}
\label{sec:data_explore}

We first explore and characterize the properties of the HSI data. Our dataset consists of five patients who have undergone glioma resection; for each patient three to four HSI images were acquired, for a total of 18 images.
The spatial resolution of each image is $1600 \times 1600$ pixels, and it contains 104 spectral channels in the range $ 468 - 787 $ nm.
Due to the implemented video adapter, only a circular central section of the focal plane is illuminated, hence the edges and corners carry no information. Additionally, due to light fall off of the scene illumination and the coupling optics the signal-to-noise ratio decreases towards the edges, with many missing pixels in that region. Finally, we observe areas of saturated pixels due to reflection.

\begin{figure}[t]
    \centering
    \includegraphics[width=\textwidth]{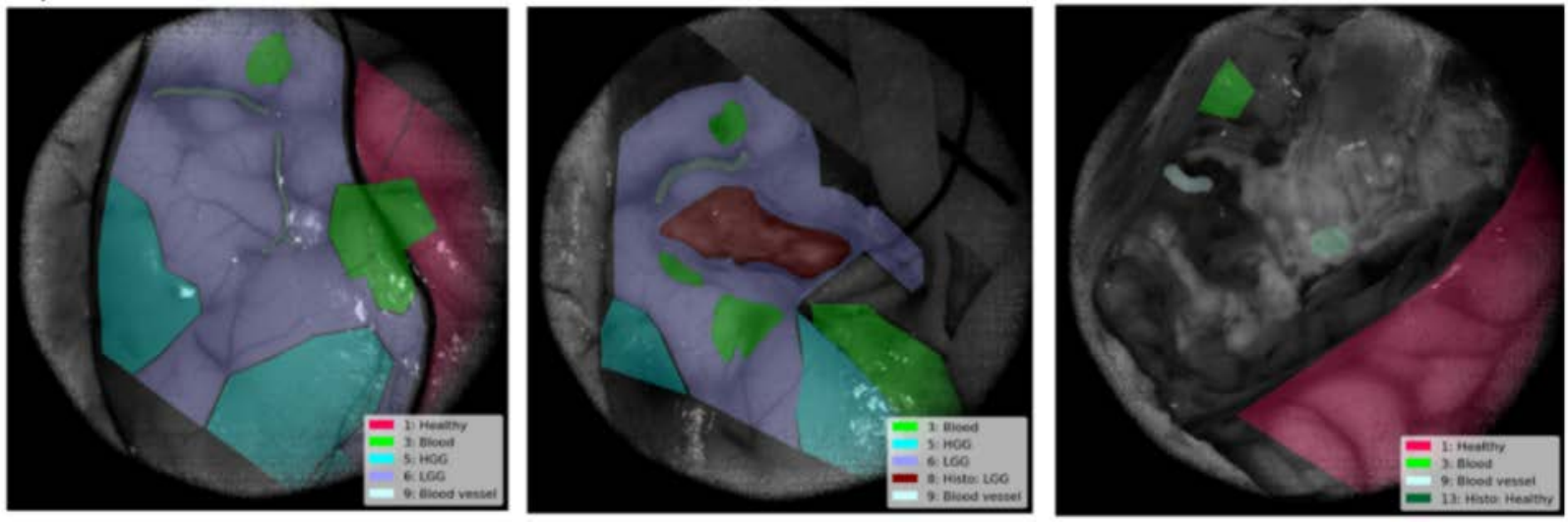}
    \caption{Three annotated images from one patient. The grayscale image represents intensity in the 660 nm channel, and the color overlays describe annotations by the surgeon. Areas without colors were not assigned to any class and belong therefore to the background class. The three images were taken before, during, and at the end of the resection process respectively.}
    \label{fig:data_pat1}
\end{figure}

\begin{figure}[t]
    \centering
    \includegraphics[trim={0 0 0 0.75cm},clip,width=0.5\textwidth]{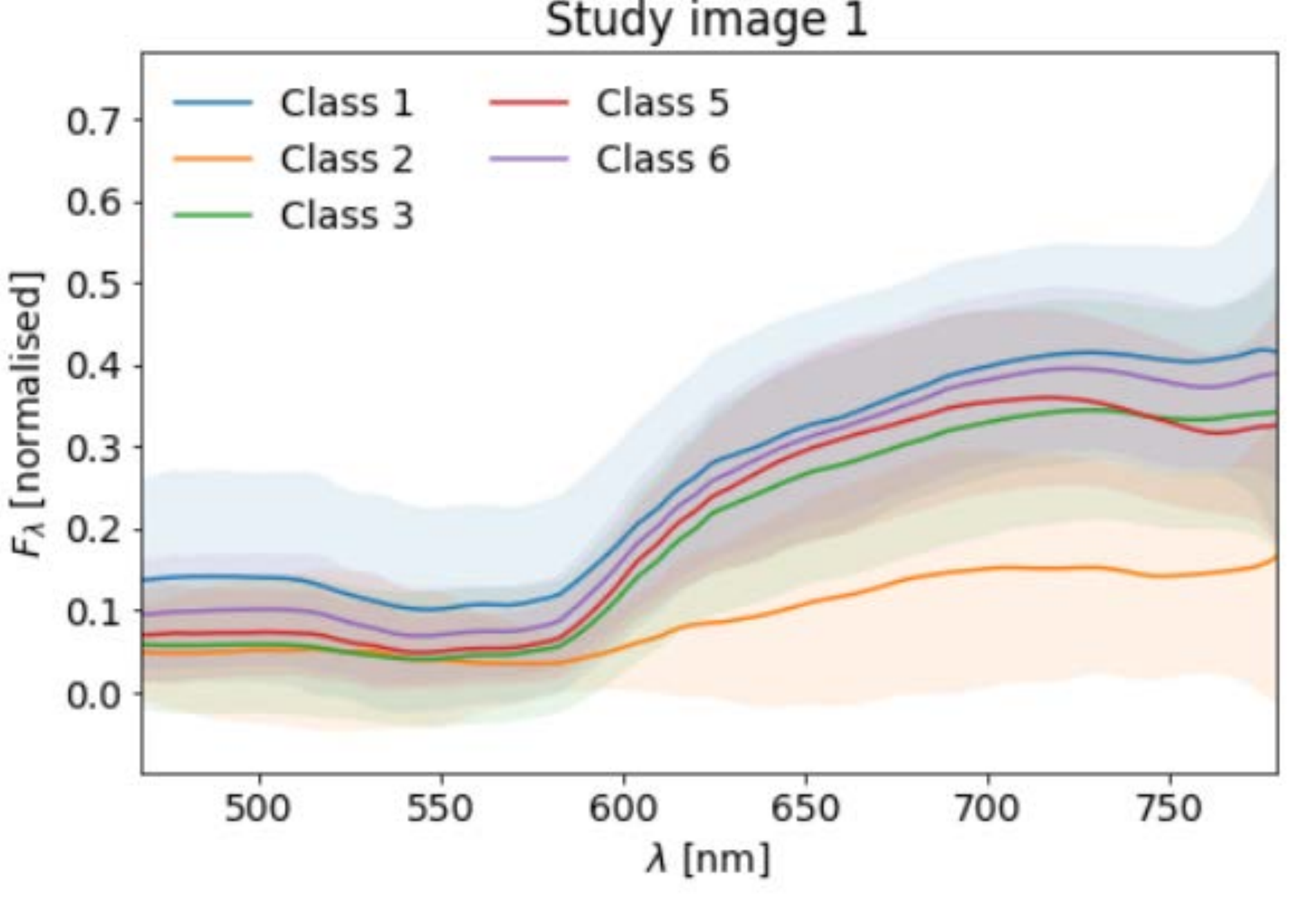}
    \caption{Normalized spectra (reflective flux) of all pixels belonging to annotated regions of one HSI image (study image 1). Only classes present in this image are shown: (1) Healthy;
    (2) Foreign object;
    (3) Blood;
    (5) HGG;
    (6) LGG. The central solid lines represent the mean spectra of all pixels belonging to each class, and the shaded areas describe the $1 \sigma$ region. 
    }
    \label{fig:spec_pix}
\end{figure}

The following classes were annotated by the operating surgeons:
\begin{enumerate*}[label=(\arabic*)]
\setcounter{enumi}{-1}
    \item Background: area to be ignored;
    \item Healthy: healthy tissue;
    \item Foreign object: surgical tools, markers etc.;
    \item Blood: Area occluded by liquid blood;
    \item Coagulation: Area occluded by coagulated blood;
    \item HGG: High-grade glioma;
    \item LGG: Low-grade glioma;
    \item Histo HGG: Histologically confirmed HGG;
    \item Histo LGG: Histologically confirmed LGG;
    \item Blood vessel: Capillaries;
    \item White matter: Typically deeper brain section exposed after surgery;
    \item Deep Cortex: Deeper cortex area, also exposed after surgery;
    \item Pia: Pia mater, the innermost of the meninges;
    \item Histo Healthy: Histologically confirmed healthy tissue.
\end{enumerate*}
In this study, we limit our classification analysis to the healthy and LGG classes (aggregating both histologically confirmed areas and those that are not).

We show in Fig.~\ref{fig:data_pat1} three exemplary annotated images from one patient, taken before, during, and after the resection operation.

In order to assess the distinguishing power of our data, we first investigate the pixel-level distribution of the measured spectra, which we show for one image in Fig.~\ref{fig:spec_pix}.
Here we see qualitatively that the spectra of most classes have largely overlapping distributions, with the possible exception of class 2 (``Foreign object"). To quantify this, we define the spectral angle mapper (SAM) distance between two spectra $s_j, s_j$ as \cite{kruse1992spectral}

\begin{equation}
\mathrm{SAM} (s_i, s_j) = \cos^{-1} \left( \frac {\sum_{l=1}^N s_{il} s_{jl}} {\left[ \sum_{l=1}^N s_{il}^2  \right]^{1/2}  \left[ \sum_{l=1}^N s_{jl}^2  \right]^{1/2}} \right) \, ,
\end{equation}
where $N$ is the number of spectral channels. We then use this distance to calculate the intra- and inter-cluster distances of the spectra, where we assume each cluster to be the set of pixels belonging to a given tissue class. The quantitative analysis confirms the qualitative impression: in all cases the intra-cluster distances are larger than the inter-cluster centroid distances, i.e. the clusters are largely overlapping and not well separated. A classical analysis with a $\chi^2$ test returns a similar result: the null hypothesis that all pixels are drawn from the same spectral distribution is not falsifiable, except for some images, at weak significance, for the ``Foreign object" class (the smallest $p$-value is found to be $p = 0.1$ between the classes ``Foreign object" and ``HGG").

\begin{figure}[t]
    \centering
    \includegraphics[width=\textwidth]{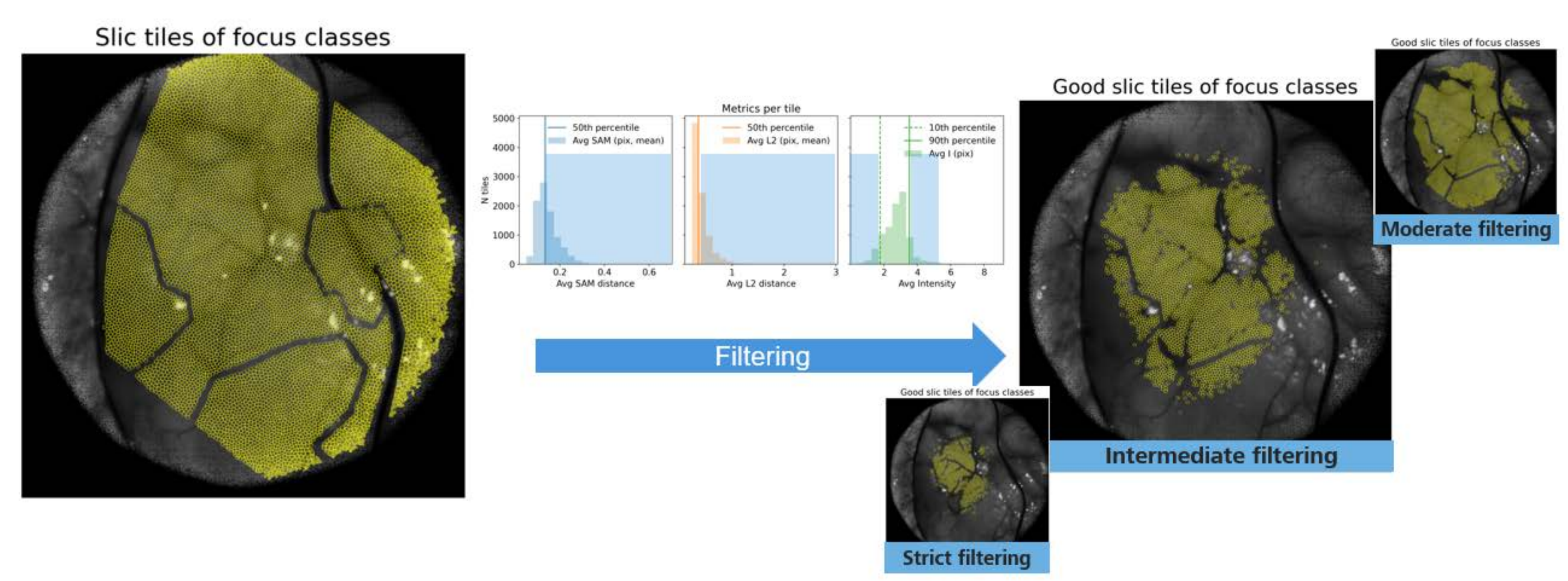}
    \caption{Creation of SLIC tiles. The left panel shows the obtained boundaries of generated SLIC tiles for one HSI image, where the tiles straddling annotation class boundaries were discarded. The central histograms show the tile filtration strategy: only tiles with high spectral uniformity are kept, following cuts on the SAM and L2 distances distributions, and tiles that are especially dark or bright are discarded, following cuts in the intensity distribution. The right panels show the resulting ``good tiles" sets that can be obtained depending on the filtration level.}
    \label{fig:slic_method}
\end{figure}

In order to improve signal-to-noise and increase the discriminating power of the data, we thus aggregate pixels by defining macro-pixels (tiles) based on the SLIC algorithm \cite{achanta2010slic}. The algorithm creates tiles by aggregating a given number of neighboring pixels in such a way as to minimize a chosen distance between them, in this case set to be the SAM distance. Instead of considering the spectra of individual pixel, we can now compare the distributions of spectra at the tile level. Furthermore, the signal-to-noise can be boosted by selecting a subset of tiles with good uniformity, and excluding outliers in brightness (particularly dark or saturated regions), as shown in Fig.~\ref{fig:slic_method}.

\begin{figure}[b]
    \centering
    \includegraphics[trim={0 0 0 0.5cm},clip,width=0.5\textwidth]{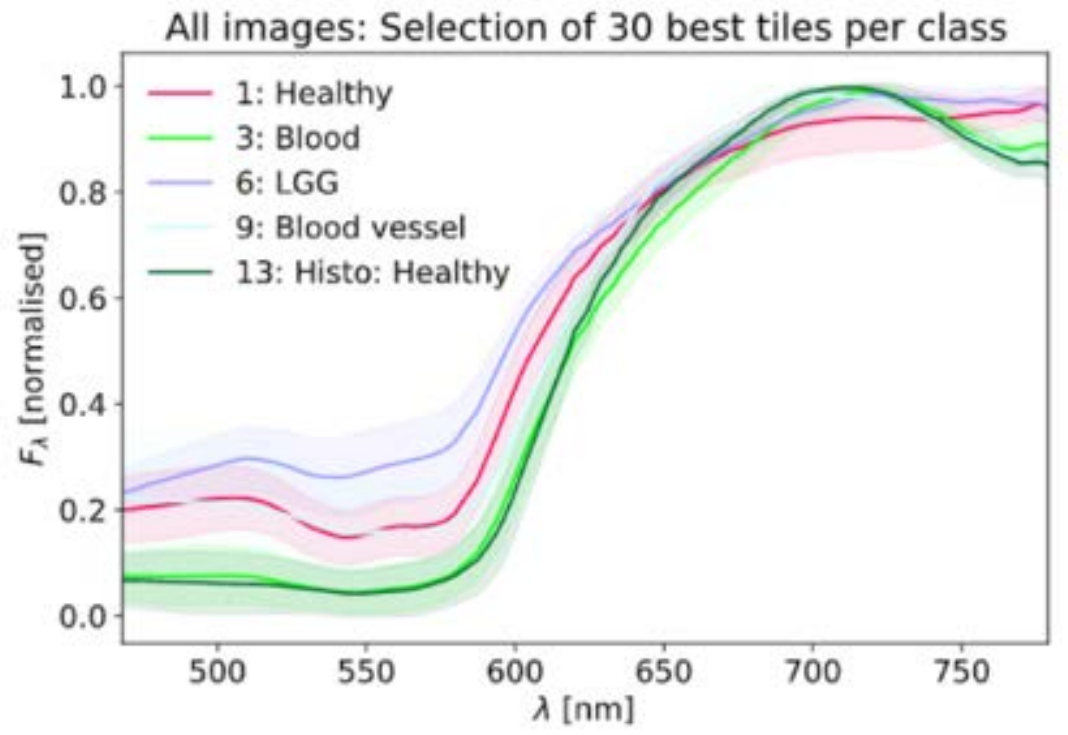}
    \caption{Normalized spectra (reflective flux) of the ``good" tiles belonging to annotated regions of one HSI image. Only classes present in this image are shown. The central solid lines represent the mean spectra of all pixels belonging to each class, and the shaded areas describe the $1 \sigma$ region. We can see that the differences among classes become statistically significant.}
    \label{fig:spec_tile}
\end{figure}

By using this procedure, we manage to obtain distributions that are statistically separable between the tissue classes, as can be seen in Fig.~\ref{fig:spec_tile}.
The class separations can achieve high statistical significance (down to a $p$-value of 0) for aggressive tile filterings, at the cost however of retaining a much reduced number of tiles.
This result is comparable with the findings by Ref.~\citeonline{fabelo2019deep}.

The results of the spectral analysis indicate that some differences between the tissue classes are visible in their spectra, albeit at low statistical significance, and are therefore encouraging: next we want to test whether, by using classical machine learning or deep learning techniques, an algorithm can be found that can classify the individual image regions to a sufficient degree of accuracy.

\subsection{Classical methods for tissue differentiation}
\label{sec:classical_methods}

Based on the results of Sec.~\ref{sec:data_explore}, we want to first investigate whether classical machine learning (ML) methods can be used to distinguish cancerous from healthy tissue. Here we focus on the classification of tiles belonging to our focus classes of ``healthy" and ``LGG", aggregating in both cases tiles that are histologically confirmed and those that are not.

We pre-process the data by creating SLIC tiles as described in Sec.~\ref{sec:data_explore} above, with a target size of 200 pixels per tile, considering tiles consisting entirely of ``healthy" or ``LGG" pixels, and applying the following quality selection criteria to the tiles: i) lowest 50\% percentile in average SAM distance among the tile pixels; ii)  lowest 50\% percentile in average L2 distance among pixels spectra; iii) average intensity $\bar I$ of the tile pixels should be in the percentile range $10\% < \bar I < 90\%$. These quality cuts restrict the dataset to particularly homogeneous tiles, and exclude in particular blood vessels, underexposed and oversaturated areas.

From the complete dataset of 5 patients and 18 images we thus obtain $ 8671 $ tiles. Given the limited number of patients at this stage, we define training and test sets by randomly splitting all tiles into a training and a test set ($ 6620 $ training tiles and $ 2051 $ test tiles). Each tile is then isolated, spatially padded with zeros to a shape of $40 \times 40 \times 104 $, and saved to disk.

We then choose three classical ML methods: a Random Forest (RF) classifier, a Support Vector Machine (SVM), and a Multi-Layer Perceptron (MLP), which we train on the training tiles and evaluate on the testing tiles. In each case, we consider as input features of the models the tile spectral information averaged over all pixels in each tile. Thus, spatial information within the tile is disregarded by the classical methods.

We report the results of this approach in Sec.~\ref{sec:res_classical} below.

\subsection{Deep Learning methods  for tissue differentiation}
\label{sec:dl_methods}

Deep convolutional neural networks (CNNs) are often used for classification tasks in computer vision. CNNs are the perfect tool to extract the most relevant information from the input data and use it to solve the desired problem. The main drawback of this approach is the need for a large and diverse training dataset that represents well the process that the CNN should approximate.

In the medical domain, the dataset acquisition is often a challenging task that requires time and huge amount of human effort. Thus, the majority of the datasets are small, and the deep learning solution should account for the data deficiency. As discussed in Sec.~\ref{sec:data_explore}, to create a suitable dataset for training, we use tiles instead of the full hyperspectral images. Thus, we end up with approximately $\sim 5000$ training examples instead of only 10-14 images in the dataset.

\begin{figure}
    \centering
    \includegraphics[trim={0 0 0 0cm},clip,width=0.75\textwidth]{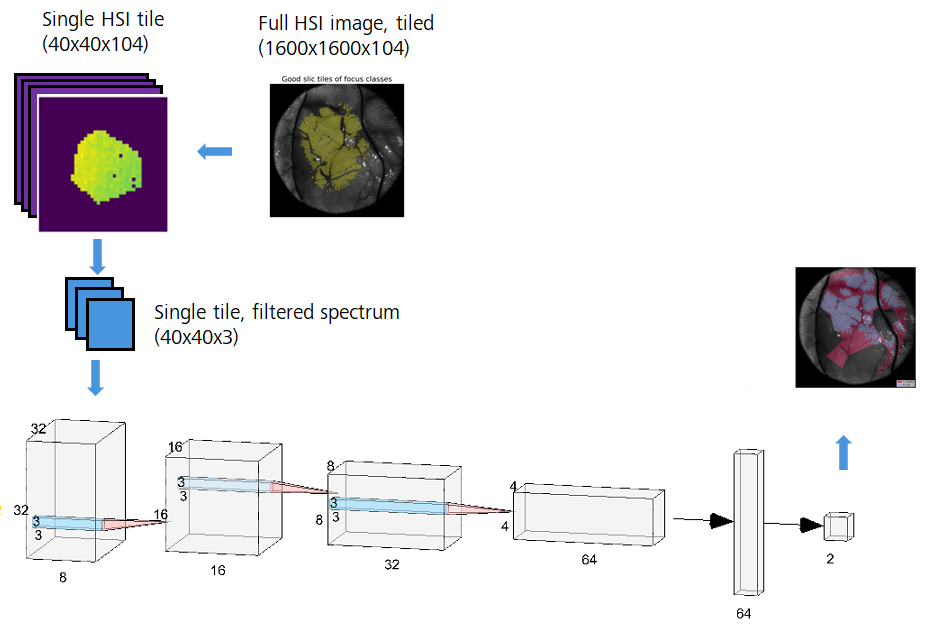}
    \caption{The proposed deep learning workflow. First, the input hyperspectral image is divided into SLIC tiles, where each tile shares similar spectra and intensities. Every tile is passed through the deep convolutional neural network that assigns healthy or tumor class to it. After all relevant tiles are processed by the neural network, the resulting prediction shows tumor and healthy areas in the image. For visualization, we show only one spectral channel out of 104, thus images appear in grayscale. The network architecture is shown in the bottom part. Here each encoder block consists of a convolutional layer with feature dimensions written at the bottom of the block and kernel size ($x,y$) shown near the blue rectangle. The spatial dimensions ($x,y$) of the input tensor are shown on the top and left vertices of each convolutional block. After each block, we also apply MaxPooling and BatchNorm layers. }
    \label{fig:cnn_arhitecture}
\end{figure}

We use a CNN, see Fig.~\ref{fig:cnn_arhitecture} for the overview of the network architecture, that takes tiles as an input and predicts tissue class as an output. Similarly to the classical methods from  Sec.~\ref{sec:classical_methods}, we consider only two classes: healthy and LGG. The encoder path consists of several convolutional layers, followed up by pooling operations; at the readout layer network features are compressed to a classification vector with class probability values in the $[0, 1]$ range. The highest value indicates the predicted class. Due to the strong correlation between neighboring spectral channels, we aim to reduce the number of spectral channels needed for the prediction. Thus, we design the neural network architecture such that the first layer reduces the number of features from 104 to 3, 6 or 12 by forming meta-channels, which are learnable linear combinations of the original input channels. Then the architecture follows the standard encoder structure. 

The evaluation of the trained model is performed with two strategies: testing and inference. The testing is the classical method to evaluate deep learning solutions, where some portion of the dataset (in our case, SLIC tiles) is left for testing and that data is not seen during training. During inference, the SLIC tiles are instead re-created from the full hyperspectral image, then  the CNN is applied to each tile independently. Thus, the prediction for the whole hyperspectral image can be obtained. In this case all tiles are used to create the prediction, even those that are of poor quality. This evaluation method helps to understand how the neural network would perform in practice, when applied during surgery.

\subsection{Spectral channels selection }
\label{sec:channel_selection}

\begin{figure}
    \centering
    \includegraphics[trim={0 0 0 0cm},clip,width=1.0\textwidth]{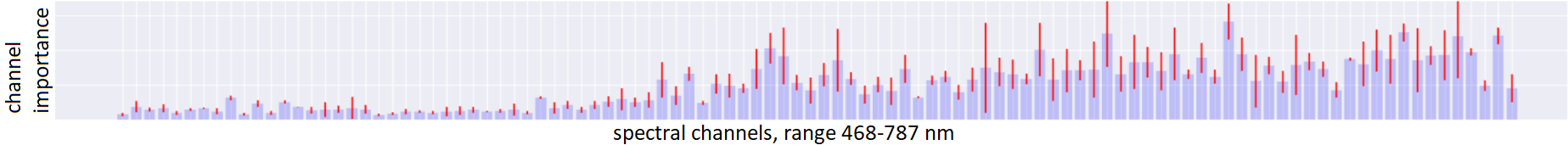}
    \caption{Explainability analysis of the ensemble of the neural networks. For each spectral channel on the $x$ axis the channel importance is calculated and averaged though the ensemble. The higher the bar, the more important is the channel for the tissue classification task. The red line indicates the standard deviation of the channel importance over the ensemble.}
    \label{fig:channel_selection}
\end{figure}

The main limitation in this exploratory feasibility study is the acquisition speed. Spectrally resolving detector arrays, as used in this study, can acquire HSI data at video rate. Here, we however chose to trade acquisition speed for spectral resolution by using a snapscan camera design. In the snapscan camera, spectral filters are deposited row-wise across the image sensor requiring spatial scanning of the scene for HSI data acquisition. However, by limiting the number of spectral channels and instead depositing the spectral filters in a mosaic pattern across the sensor, spatial scanning can be circumvented, and snapshot spectral imaging achieved. With design modifications, the HSI measurement technique can be straightforwardly translated to a real-time setting, achieving frame rates of 15-30 fps as required in the clinical setting. In order to apply our method in the OR, we thus focus our efforts towards selecting the most relevant spectral channels for tissue differentiation. By doing that, we can decrease the acquisition time and create a solution that can be later used in real-time applications.

We thus search for the 3, 6, and 12 most important spectral channels that the neural network needs for the prediction. In order to select those channels, we use the information from an ensemble of trained neural networks. We investigate for each network what channels contributed the most to its prediction, given that the network achieves good results (above 80$\%$ accuracy) on the test dataset. Then we accumulate the channel importance scores from the individual networks and select the most important spectral channels for the final CNN input, see Fig. \ref{fig:channel_selection} for an illustration of the channel importance. Note, that after the channel selection, the neural network is retrained with only important channels.

To find what channels networks favor the most we apply the state-of-the-art GradientShap explainability method~\cite{ancona2018towards}. However, there are other techniques in the machine learning community, such as IntegratedGradients or Occlusion-based Attribution~\cite{ancona2018towards}. The output of these methods is an attribution score to each input feature (the spectral channels) concerning the first target output. Positive attribution scores indicate a positive contribution of the input at that specific position to the final prediction, whereas negative scores indicate the opposite. The strength of the contribution is indicated by the attribution score’s size.

\subsection{Reliability and out-of-distribution predictions using ensemble models
}
\label{sec:ensemble}

It is well known that a single neural network can be overconfident \cite{ensemble}. The predictions can have very high score and be incorrect. This situation often occurs when input data is no longer statistically similar to the training dataset. Since the neural network was never trained with such an out-of-distribution sample, it still predicts one of the classes that it was trained for.

The state-of-the-art approach to estimate how confident is the network about its prediction is to use an ensemble of neural networks. The final prediction is computed based on the average class probability prediction of multiple neural networks that have similar architecture and are trained on the same dataset. The assumption is that for the out-of-distribution sample networks in the ensemble would not agree on the predicted class, thus the average prediction scores (class probabilities) would be low for this sample. A simple thresholding technique can rule out low score predictions and mark the corresponding input samples as ``unknown''. The predictions where all networks agree are considered reliable. We illustrate the effect of using the ensemble method to verify the prediction quality and identify the out-of-distribution areas in the input hyperspectral image.

\section{Results}
\label{sec:results}

We performed the data analysis and model training and evaluation on a Linux workstation with 12  dual-core CPUs with a total 64 GB RAM, and an Nvidia GeForce RTX 3090 GPU, with 24 GB RAM.

\subsection{Results for classical methods}
\label{sec:res_classical}

\begin{figure}[b]
    \centering
    \includegraphics[trim={0 0 0 0cm},clip,width=0.75\textwidth]{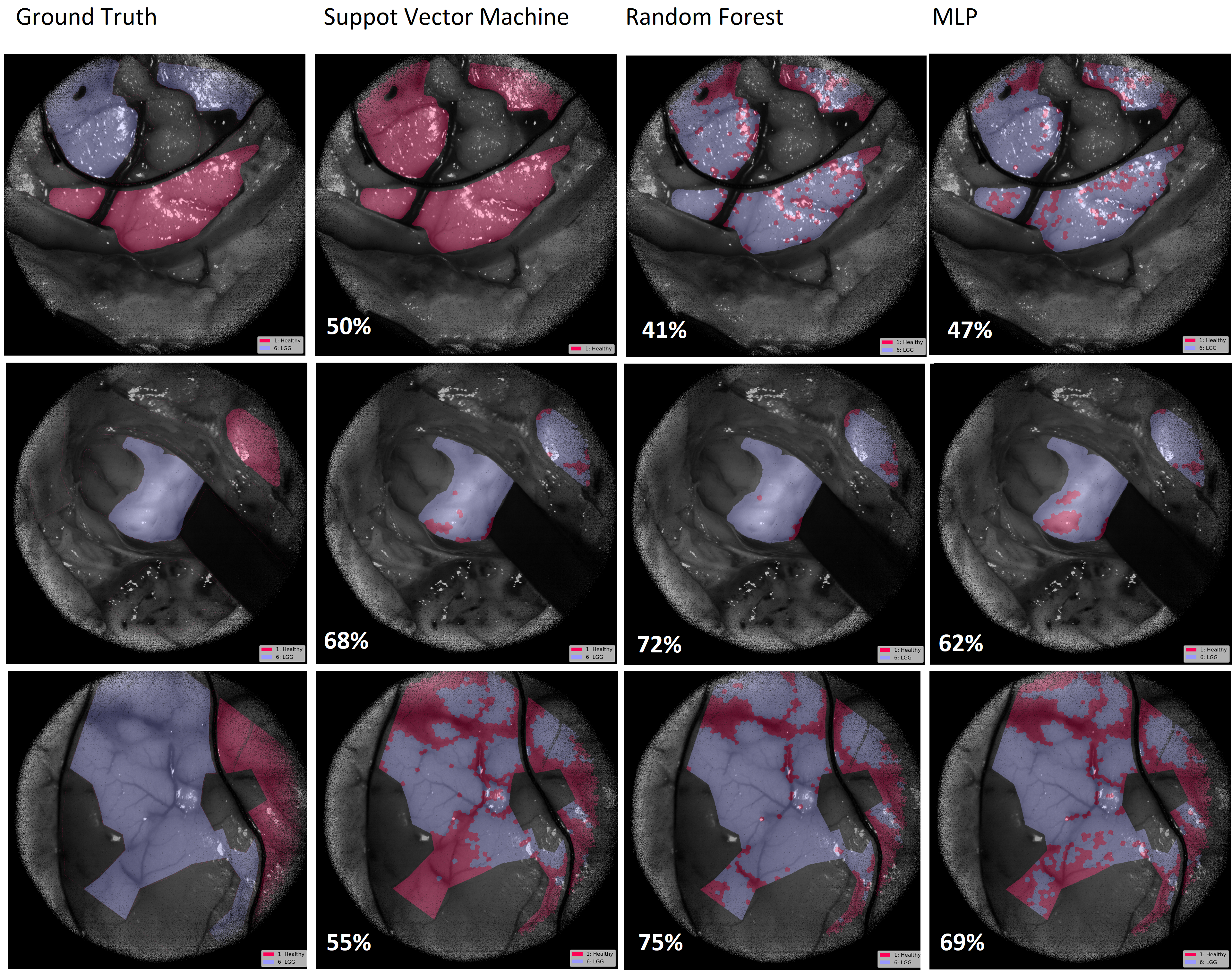}
    \caption{Inference on a selection of three full images using the classical ML methods. The left panels show the ground-truth labels for the two focus classes (healthy in red and LGG in blue), while the following panels show the inference results for the Support Vector Machine, Random Forest and Multi-Layer Perceptron methods respectively. The prediction quality is significantly lower than on the good-tiles test set.}
    \label{fig:ML_inference}
\end{figure}

We first report the results of evaluating the classification models on the test set of unseen tiles. This is the default choice, but it still contains only ``good" tiles that belong to the two classes we consider, i.e. it does not include any out-of-distribution tissue, nor any lower-quality tiles.
All classical models underwent hyperparameter tuning and the results are referred to a 0.5 operating point.

Our first baseline classical ML model, a Random Forest classifier with 100 trees, achieves an overall 86\% accuracy on the test set, with 94\% precision and 88\% recall.

The second model considered, an SVM with Radial Basis Function (RBF) kernel, achieves a 91\% accuracy (95\% precision, 92\% recall).

Finally the third model, an MLP classifier with one hidden layer, achieves an improved accuracy of 92\%, with 98\% precision and 91\% recall.

While promising, these results are based on the selected good-quality tiles only. In order to assess a more realistic performance in the operating theatre, we apply the trained models in inference mode onto the full images. With this procedure, we achieve as expected a lower accuracy, which is typically between $\sim 50 \%$ and $\sim 70\%$, as we show for three example images in Fig.~\ref{fig:ML_inference}. From the figure it is qualitatively clear that these classical classifiers often produce incorrect predictions over broad image regions.

Note also that this evaluation method does include all lower-quality tiles that were excluded from our dataset as prepared in Sec.~\ref{sec:classical_methods}, but it also includes the tiles from the training set, and thus it still represents an optimistic estimate of the actual model performance in the wild.

\subsection{Results for Deep Learning methods}
\label{sec:res_deep}

 We summarize in Tab.~\ref{tab:res_dl_runs} the quantitative results of our deep learning classification models evaluated on the test dataset, where we can see that an accuracy $> 80 \%$ is generally achieved. In order to investigate the stability of the models, for each network configuration we retrain the model three times with random initialization of the trainable parameters. The ``Channel compression'' column contains the information about the first convolutional layer that compresses the 104 input spectral channels to 3, 6 or 12 channels. The other columns illustrate the performance of the neural networks, where we observe an increase in all metrics with the increase of the number of features. Thus, the more features are used, the more information can be extracted from the input spectrum. The difference between the runs with the same architecture is quite small, thus we conclude that the neural network does not depend much on the input initialization. However, the small difference in the performance supports the idea to use ensemble instead of a single network to compensate for the effect of the random initialization.

\begin{table}[htbp]
\caption{Deep learning classification models: quantitative evaluation on the test dataset.}
\label{tab:res_dl_runs}
\begin{center}
\begin{tabular}{|l|l|l|l|l|l|l|l|}
\hline
Channel compression  & Accuracy & Precision & Recall & TP   & TN  & FP & FN  \\ \hline\hline
104-3  & 76\%     & 92\%      & 74\%   & 1137 & 428 & 95 & 391 \\ \hline
104-3  & 81\%     & 93\%      & 81\%   & 1238 & 426 & 97 & 290 \\ \hline
104-3  & 78\%     & 93\%      & 77\%   & 1178 & 430 & 93 & 350 \\ \hline\hline
104-6  & 87\%     & 94\%      & 87\%   & 1334 & 444 & 79 & 194 \\ \hline
104-6  & 82\%     & 95\%      & 80\%   & 1221 & 460 & 63 & 307 \\ \hline
104-6  & 85\%     & 95\%      & 85\%   & 1298 & 450 & 73 & 230 \\ \hline\hline
104-12 & 89\%     & 94\%      & 90\%   & 1378 & 443 & 80 & 150 \\ \hline
104-12 & 86\%     & 96\%      & 85\%   & 1297 & 446 & 57 & 231 \\ \hline
104-12 & 89\%     & 94\%      & 90\%   & 1373 & 445 & 78 & 155 \\ \hline
\end{tabular}
\end{center}
\end{table}

\begin{figure}[tb]
    \centering
    \includegraphics[trim={0 0 0 0cm},clip,width=0.75\textwidth]{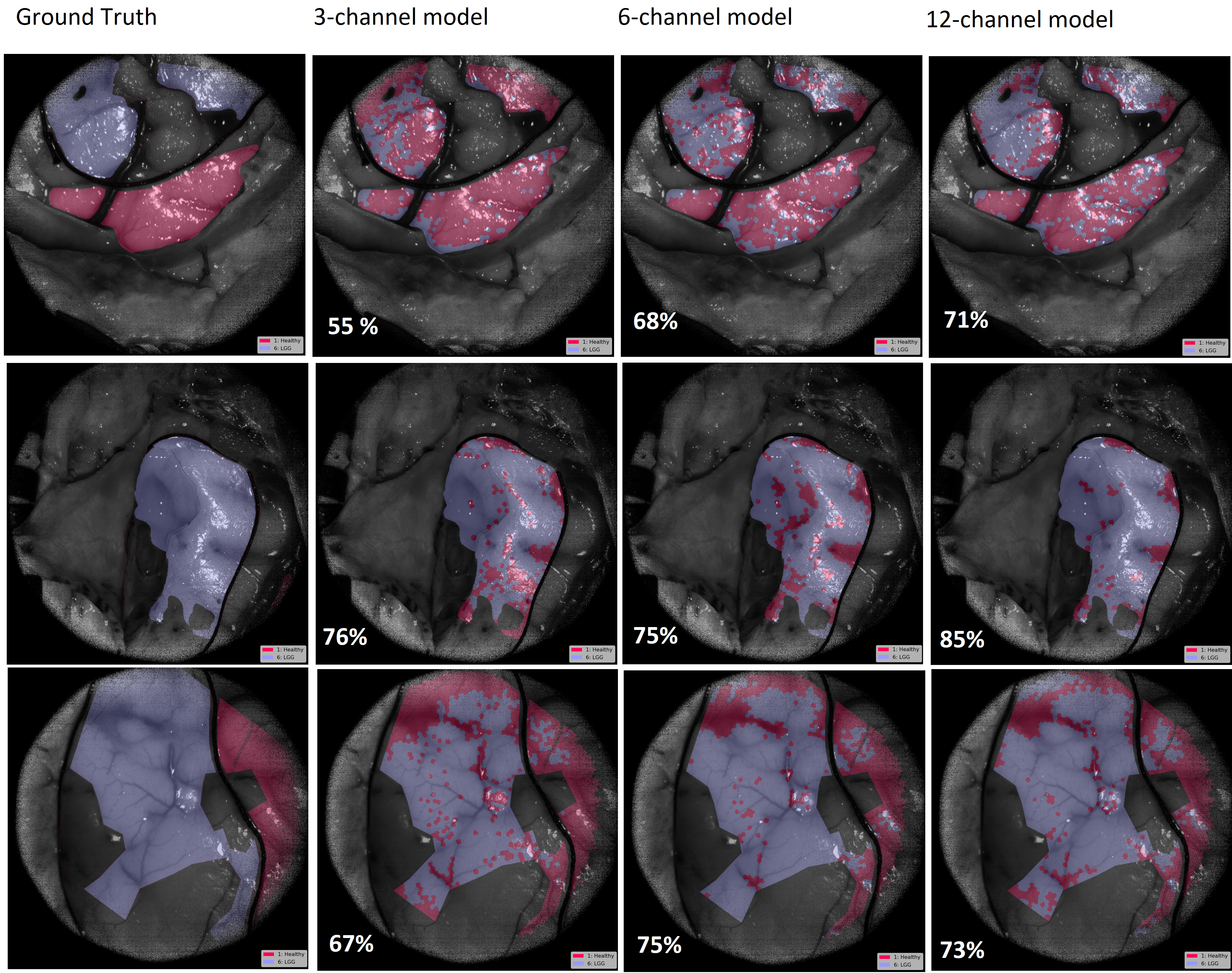}
    \caption{ Qualitative evaluation of the models that compress input spectral channels to 3, 6, and 12 meta-channels. The new channels are encoded with the deep convolutional neural network. The results illustrate the application of the trained models to input hyperspectral images taken from different patients. Healthy tissue is depicted in red, while tumor tissue is shown in blue. The first column displays annotations from the surgeon. The other columns are the results. Note, that accuracy is calculated for every image separately. The accuracy of the prediction is different from the performance of those models on the test set due to the varying acquisition quality of some areas in the images. The test dataset consists only of the good quality samples (tiles).  }
    \label{fig:qualitative_res}
\end{figure}

Figure~\ref{fig:qualitative_res} illustrates the inference results on several full images: here the models with compression to 3, 6, and 12 channels are applied to the full hyperspectral images. For some samples, the model with 3 channels under-performs compared to the models with 6 and 12 channels. This happens due to the discarding of information in the first layer of the neural network, where 104 channels were aggregated into 3. Models with 6 and 12 channels show comparable performance on the test images, both qualitatively and quantitatively.
These results are as expected qualitatively and quantitatively superior to those obtained with classical ML methods shown in Fig.~\ref{fig:ML_inference}, not only due to the superior algorithm used, but also as for the deep learning case the full spectral information of every pixel of each tile is retained. When using classical ML techniques, only the average spectrum over all pixels was considered as the descriptive feature of each tile.

\begin{figure}[t]
    \centering
    \includegraphics[trim={0 0 0 0cm},clip,width=0.9\textwidth]{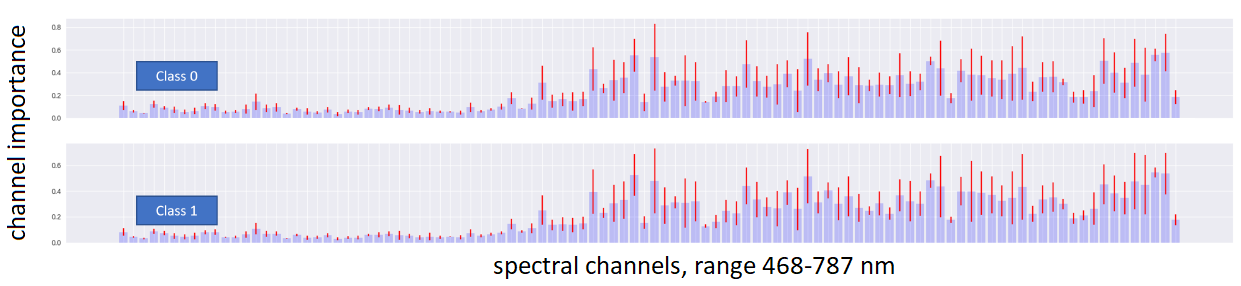}
    \caption{Importance of the spectral channels. The blue bars correspond to the mean importance score, while the red lines illustrate the standard deviation across 3 models. The models have exactly the same architecture, but trained with random initialization, thus differ in the initial values of all trainable parameters. }
    \label{fig:channels_12}
\end{figure}

Then we study the channel importance, where we identify the most important channels from the trained models. We illustrate the explainability analysis on the model with the compression to 12 channels. First, we show the channel importance in Fig.~\ref{fig:channels_12}, where the most important channels are concentrated in the red part of the spectrum, starting from approximately 650 nm. Then we select the 12 channels with the highest scores and retrain the model using these channels only. Results on the test dataset show that the model achieves accuracy of 81$\%$, precision of 94 $\%$, and recall 80 $\%$. Compared to the results from Tab.~\ref{tab:res_dl_runs}, we conclude that the model performance is marginally lower compared to the network that was trained on the full spectrum with later channel aggregation. However, the results still show that a channel reduction is possible, but more research is required in order to identify an optimal channel selection. 

\begin{figure}
    \centering
    \includegraphics[trim={0 0 0 0cm},clip,width=0.75\textwidth]{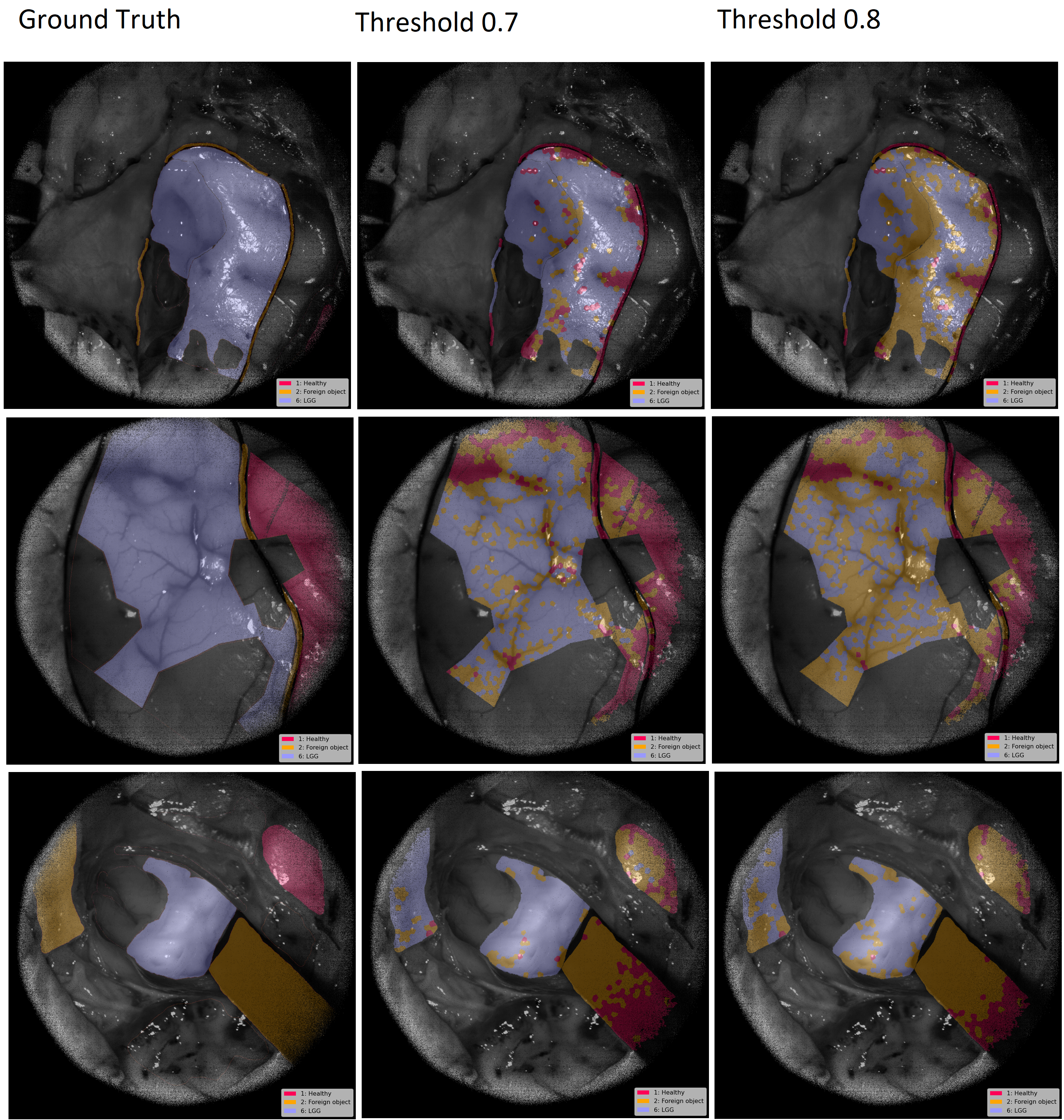}
    \caption{ Results of the applying ensemble to the full hyperspectral images. The first column represents the ground truth, where the healthy areas are marked in red, tumor areas are marked in blue and unknown regions are orange. The second column illustrates the results after thresholding the prediction scores at 0.7 and the third column shows the results with a 0.8 threshold.  }
    \label{fig:ensemble}
\end{figure}

The final evaluation step is the uncertainty estimation via ensemble networks. As discussed in Sec.~\ref{sec:ensemble}, instead of training one model, we train several models and aggregate their predictions. All neural networks are trained to predict one of the two classes: tumor or healthy. By using the ensemble, we can threshold the average score from all trained models and add the `unknown' class. All predictions that are below the threshold will be marked as `unknown'. We trained 10 models with random initialization and illustrate the ensemble results with two thresholds 0.7 and 0.8, see Fig.~\ref{fig:ensemble}. We added some areas from other classes, which the networks never saw during training. We expect that all those areas should be marked as `unknown'. According to the qualitative results from Fig.~\ref{fig:ensemble}, most of the unknown areas are already found when thresholding the predictions with 0.7. When a higher threshold is applied, more areas fall into the `unknown' category. The desired balance between the amount of predicted area and the reliability of this prediction can be achieved by  adjusting the threshold.

\section{Conclusions}
\label{sec:conclusions}

In this study, we illustrated the development of a deep-learning algorithm to visualize LGG based on snapscan HSI technology. RGB and HSI images were acquired on 5 patients with LGG intra-operatively and were annotated by the operating surgeon for subsequent training. 

The results using classical methods including Random Forest, SVM with RBF and MLP classifiers show an accuracy $> 90\%$ using only ``good'' tiles (i.e. tiles with low spectral variability and excluding under- and over-exposed regions) and all 104 channels. When using all tiles, a much lower accuracy was found using these classical methods, which is typically $\sim 70\%$ at best. The accuracy using all tiles was increased to $> 80\%$ using deep learning methods. Furthermore, the performed analysis of channel importance demonstrated that the channels in red and NIR, i.e. from ca. 650 nm to 780 nm, carry most classification weight. A channel reduction experiment showed that the overall performance does not decrease dramatically with channels being reduced to either 12 or 6 channels. A much worse performance was only found when the number of channels was reduced to 3. 

The presented work is still ongoing in cooperation with UZ Leuven and IMEC and has some limitations: (i) Only 5 patients with altogether 18 study images were acquired so far and this dataset might still be too small to address inter-patient variability. Indeed, due to the limited number of patients, we have built our training and testing datasets randomly from all patients.
(ii) Only tiles inside the annotated areas were evaluated using the deep-learning method.
Therefore, in the near future, we will carry out additional research as we continue our data acquisition in UZ Leuven. First, once more data will be available,
we look forward to extend the train/test split to the case of datasets that are separated patient-wise, in order
to validate the trained network on unseen data and to verify whether it starts to generalize on larger datasets. Moreover, we will validate the trained networks on the full images, including all unannotated areas, as a next step towards a more practical tool that can aid surgeons in the OR resecting LGGs. Despite these facts, the findings of this preliminary work demonstrate the feasibility of applying HSI technology to differentiate LGG from healthy tissues, which paves the way towards a real product that could be used in the OR in the near future.

\appendix


\bibliography{report} 
\bibliographystyle{spiebib} 

\end{document}